\documentclass[aps,prl,reprint,preprintnumbers,showpacs,floatfix,nofootinbib,superscript address,longbibliography]{revtex4-2}
\usepackage[utf8]{inputenc}
\usepackage{amssymb}
\usepackage{hhline}
\usepackage{amsmath}
\usepackage{mathtools}
\usepackage[dvipsnames]{xcolor}
\usepackage{multirow,tabularx}
\usepackage{graphicx}
\usepackage{natbib}

\usepackage{xspace}
\usepackage{xstring}
\usepackage{titlesec}
\usepackage{parskip}
\allowdisplaybreaks
\usepackage{braket}

\newrobustcmd{\pea}[1]{%
	\emph{#1}\textbf{\ \ \ ---}
}
\titleformat{\paragraph}[runin]{\normalfont\normalsize\bfseries}{\emph\theparagraph}{1em}{\pea}

\newcommand*{\eg}{e.g.\@\xspace}
\newcommand*{\cf}{c.f.\@\xspace}
\newcommand*{\fig}{fig.\@\xspace}
\newcommand*{\eq}{eq.\@\xspace}
\newcommand*{\eqs}{eqs.\@\xspace}

\newcommand*{\lhs}{l.h.s.\@\xspace}

\usepackage{hyperref}
\hypersetup{%
     colorlinks = true,%
     linkcolor = Blue,%
     citecolor = Blue,%
     filecolor = Blue,%
     urlcolor = Blue%
     }%

\begin{document}

\title{No Dark Matter Axion During Minimal Higgs Inflation}

\author{Claire Rigouzzo}
\email{claire.rigouzzo@kcl.ac.uk}
\affiliation{Laboratory for Theoretical Particle Physics and Cosmology,\\
	King's College London, London, United Kingdom}
\author{Sebastian Zell}
\email{sebastian.zell@lmu.de}
\affiliation{Arnold Sommerfeld Center, Ludwig-Maximilians-Universit\"at, Theresienstraße 37, 80333 M\"unchen, Germany}
\affiliation{Max-Planck-Institut für Physik, Boltzmannstr. 8, 85748 Garching b.\ M\"unchen, Germany}


\begin{abstract}
We study minimal versions of Higgs inflation in the presence of a massless QCD axion. While the inflationary energy scale of the metric variant is too high to accommodate isocurvature bounds, it was argued that Palatini Higgs inflation could evade these constraints. We show, however, that an energy-dependent decay constant enhances isocurvature perturbations, implying that axions can at most constitute a tiny fraction $< 10^{-5}$ of dark matter. This conclusion can be avoided in Einstein-Cartan gravity by an additional coupling of the axion to torsion, albeit for a very specific choice of parameters. Analogous constraints as well as the possibility to alleviate them are relevant for all inflationary models with a non-minimal coupling to gravity, including Starobinsky inflation and certain classes of attractor models.
\end{abstract}

\maketitle

\paragraph*{Inflation \& dark matter}
In our Universe, about 85\% of matter has only been observed through gravitational effects. Understanding the microscopic nature of this dark matter (DM) remains one of the greatest challenges in cosmology. Another fundamental mystery is the origin of the initial conditions for the hot Big Bang.  The leading explanation is  cosmic inflation \cite{Starobinsky:1980te, Guth:1980zm, Linde:1981mu, Mukhanov:1981xt} -- an early phase of accelerated expansion -- which successfully accounts for the near-homogeneity and isotropy of the Universe and is strongly supported by precision measurements of the cosmic microwave background (CMB) \cite{Planck:2018jri, BICEP:2021xfz}. 
A wide range of DM models have been developed (see \cite{Bertone:2016nfn}). Similarly, a plethora of
inflationary scenarios have been put forward (see \cite{Martin:2013tda}). The key difficulty lies in distinguishing viable proposals from the multitude of possibilities, with the ultimate goal of identifying the mechanisms that govern our Universe. In this work, we reveal an incompatibility between two leading candidates: QCD axions \cite{Peccei:1977hh,Weinberg:1977ma,Wilczek:1977pj} and Higgs inflation (HI) \cite{Bezrukov:2007ep}.

Originally, axions were introduced to address the strong CP-problem of QCD, originating from the experimental fact that strong interactions are CP-conserving to a very good accuracy \cite{Abel:2020pzs}. This observation can be explained by promoting the CP-violating angle of QCD to a dynamical pseudoscalar field -- the axion \cite{Peccei:1977hh,Weinberg:1977ma,Wilczek:1977pj}, which does not only resolve the strong CP-problem but moreover contributes to DM (see \eg \cite{DiLuzio:2020wdo}). The QCD axion can arise, among others, from the original proposal of a spontaneously broken PQ-symmetry \cite{Peccei:1977hh,Weinberg:1977ma,Wilczek:1977pj}, from extra dimensions \cite{Witten:1984dg} (see review \cite{Reece:2024wrn}), or from local gauge invariance \cite{Dvali:2005an,Dvali:2013cpa,Dvali:2017mpy,Dvali:2022fdv}.

HI agrees excellently with CMB observations and is unique among inflationary proposals since it does not require more particles than have already been observed in experiment. HI is highly sensitive to the different equivalent formulations of General Relativity (GR) \cite{Bauer:2008zj,Rasanen:2018ihz, Raatikainen:2019qey, Langvik:2020nrs, Shaposhnikov:2020gts} (see overview in \cite{Rigouzzo:2022yan}). While the most commonly-employed metric
version suffers from a low perturbative cutoff scale \cite{Burgess:2009ea,Barbon:2009ya}, no such problems exist in Palatini HI \cite{Bauer:2010jg}, providing strong motivation for considering the Palatini variant of HI.\footnote
{Whether the low cutoff scale of metric HI, above which perturbation theory breaks down, invalidates inflationary dynamics remains debated \cite{Burgess:2009ea,Barbon:2009ya,Bezrukov:2010jz,Bezrukov:2014bra,Bezrukov:2014ipa,Barbon:2015fla,Karananas:2022byw}. It does, however, lead to large and partly uncontrollable corrections to the renormalization group (RG) running, so the high-energy values of the coupling constants cannot be computed uniquely and an uncertainty arises in inflationary predictions \cite{Bezrukov:2010jz,Bezrukov:2014bra,Bezrukov:2014ipa,Fumagalli:2016lls,Enckell:2016xse,Bezrukov:2017dyv,Shaposhnikov:2020fdv}. In addition, the reheating temperature exceeds the cutoff scale, so the end of inflation cannot be determined unambiguously \cite{Ema:2016dny} and becomes sensitive to the UV completion \cite{He:2018mgb,Bezrukov:2020txg}. No such issues arise in Palatini HI because the cutoff scale is significantly higher \cite{Bauer:2010jg}. As a result, RG running remains under perturbative control and inflationary predictions can be linked to low-energy collider measurements \cite{Shaposhnikov:2020fdv,Poisson:2023tja}, while reheating can be calculated unambiguously since the temperature remains below the cutoff \cite{Rubio:2019ypq,Dux:2022kuk}.}

A massless axion present during inflation generically sources isocurvature perturbations \cite{Turner:1990uz}, which are tightly constrained by their non-detection in the CMB. These bounds can be avoided if the axion field is absent during inflation (or reheating). This can be realized in certain UV completions, most notably those featuring a PQ symmetry, provided its spontaneous breaking occurs after inflation. In this case, however, many models are excluded due to the overproduction of topological defects \cite{Sikivie:1982qv}. In addition, the viable parameter space for post-inflationary symmetry breaking is much smaller in Palatini HI than in metric HI, owing to the lower inflationary energy scale.

\begin{figure}
	\centering
	\includegraphics[width= 0.8\linewidth]{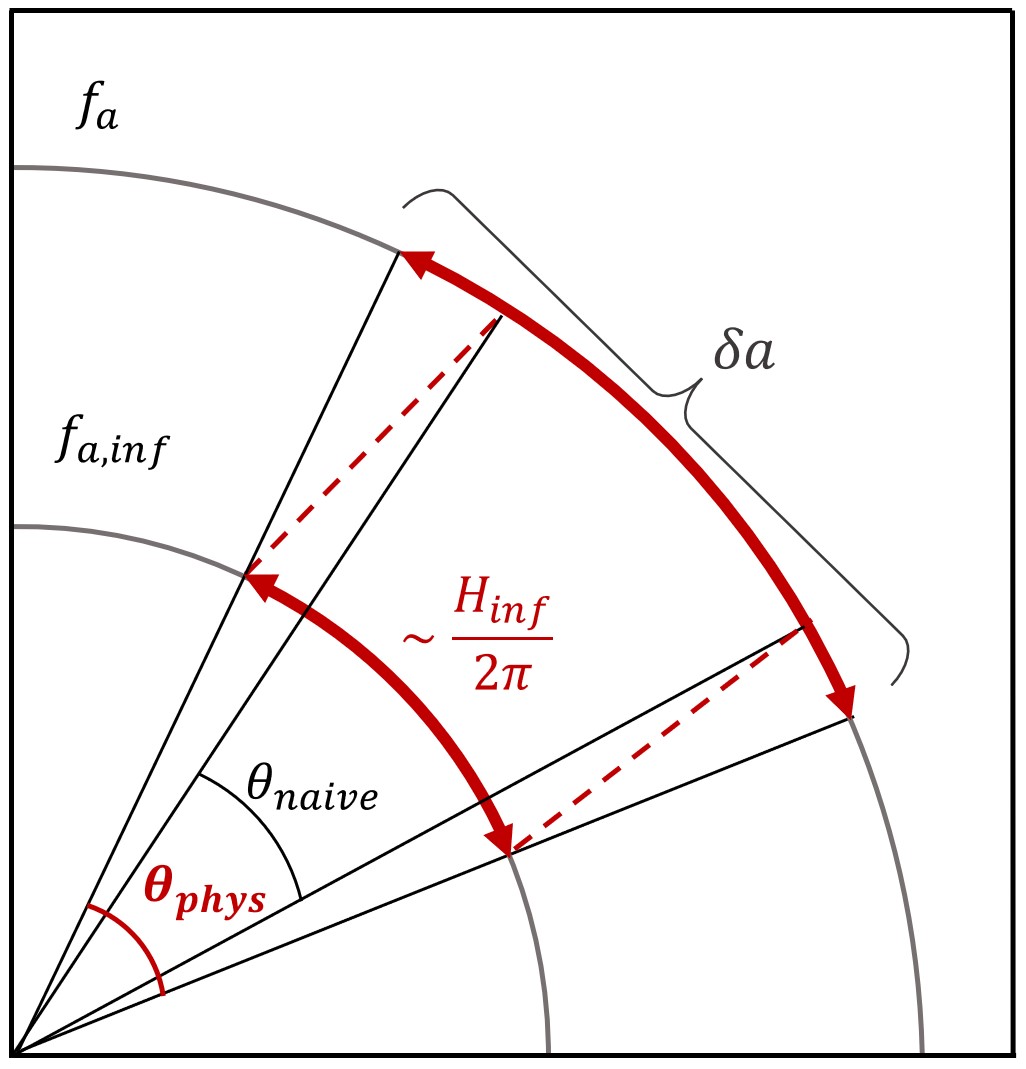}
	\caption{Representation of varying decay constant. Since the effective decay constant is reduced during inflation, isocurvature perturbations are enhanced. The dashed line corresponds to the result of \cite{Tenkanen:2019xzn}, where the change of the decay constant was not taken into account. The \textit{physical} fluctuation of the axion field is shown with $\delta a$. Figure inspired by \cite{Fairbairn:2014zta}.}
	\label{diagram}
\end{figure}

In this paper, we shall investigate the other alternative and check the compatibility of pre-inflationary axions with HI. While metric HI leads to excessive isocurvature perturbations, it was suggested that Palatini HI could tolerate
a massless QCD axion during inflation \cite{Tenkanen:2019xzn}. However, the energy dependence of the decay constant \cite{Fairbairn:2014zta, Ballesteros:2016xej} (see also \cite{Linde:1991km,Higaki:2014ooa,Choi:2014uaa,Chun:2014xva}) was not taken into account in \cite{Tenkanen:2019xzn}. Since it is smaller during inflation than in the late Universe, isocurvature perturbations are enhanced (see illustration in \fig \ref{diagram}). We point out that this is a generic feature in all inflationary models with a large non-minimal coupling to gravity. Applied to Palatini HI, we show that the presence of a massless axion field that can later act as DM is excluded. So pre-inflationary DM axions are incompatible with minimal versions of HI.

\paragraph*{Review of isocurvature bound}
CMB observations bound the magnitude of uncorrelated isocurvature perturbations $\Delta_a$ relative to the adiabatic curvature perturbations $\Delta_{\mathcal{R}}$ as $\Delta_a^2/(\Delta^2_{\mathcal{R}}+\Delta_a^2) \lesssim 0.038$ \cite{Planck:2018jri,DiLuzio:2020wdo}. Allowing for correlations between isocurvature and adiabatic mode modifies this number slightly, as discussed in Appendix \ref{app:isocurvature}, but the conclusion of the paper remains unchanged. Plugging in the measured value $\Delta^2_{\mathcal{R}}\approx 2.1 \cdot 10^{-9}$, we get \cite{Planck:2018jri}
\begin{equation} \label{isocurvatureBound}
	\Delta_a \lesssim 9.1 \cdot 10^{-6} \;.
\end{equation}
Now an axion with initial misalignment angle $\theta_i$ induces isocurvature perturbations \cite{Beltran:2006sq,Hertzberg:2008wr} 
\begin{equation} \label{isocurvatureAxion}
	\Delta_a=\mathcal{F}^a_{\text{DM}} \frac{\sigma_\theta \sqrt{2(\sigma^2_\theta+2 \theta_i^2)}}{\theta_i^2+\sigma_\theta^2} \;,
\end{equation}
where $\mathcal{F}^a_{\text{DM}}$ is the relative contribution of axions to DM and $\sigma_{\theta}$ stands for the typical quantum fluctuation of the angular axion field. Moreover, $\theta_i$ is the initial misalignment angle, which can be expressed through $\mathcal{F}^a_{\text{DM}}$ and the axionic decay constant $f_a$ (see details in appendix \ref{app:abundance} of the Supplemental Material and references \cite{RevModPhys.53.43,Borsanyi:2016ksw,Saikawa:2018rcs,Sheridan:2024vtt} therein):
\begin{equation} \label{thetaOff}
 \theta_{i} = \mathcal{F}^{a\, 1/2}_{\text{DM}} \left(\frac{1.02 \times 10^{12} \text{GeV}}{f_a}\right)^{7/12} \;.
\end{equation}

For $\sigma_\theta\gg \theta_i$, \eq \eqref{isocurvatureAxion} reduces to $\Delta_a\approx \sqrt{2} \mathcal{F}^a_{\text{DM}}$ and so the isocurvature bound can only be satisfied for a tiny $\mathcal{F}^a_{\text{DM}}$. Having a sizable contribution of axions to DM, $\mathcal{F}^a_{\text{DM}}\approx 1$, is only possible for $\sigma_\theta \ll \theta_i$, in which case \eq \eqref{isocurvatureAxion} gives $\Delta_a \approx \mathcal{F}^a_{\text{DM}} 2 \sigma_\theta/\theta_i$. In summary, the isocurvature constraint \eqref{isocurvatureAxion} implies that one of the following two conditions must be satisfied (see details in appendix \ref{app:isocurvature} of the Supplemental Material):
\begin{equation} \label{isocurvatureBoundDecayConstant}
\mathcal{F}^a_{\text{DM}} \lesssim 6.4\cdot 10^{-6} \quad\ \text{OR}\ \quad	\sigma_\theta \lesssim 4.6 \cdot  10^{-6} \frac{\theta_i}{\mathcal{F}^a_{\text{DM}}} \;.
\end{equation}
Since we are interested to have a sizable fraction of DM in axions, we shall focus on fulfilling the second condition.

\paragraph*{Argument of \cite{Tenkanen:2019xzn}}
In Palatini HI, the inflationary Hubble scale is (see \cite{Bauer:2008zj,Shaposhnikov:2020fdv,Poisson:2023tja} and derivation below)
\begin{equation} \label{Hubble}
	H_I \sim  2.6 \cdot 10^{-6} \frac{M_P}{\sqrt{\xi}} 
	\;. 
\end{equation}
Here $\xi$ sets the strength of non-minimal coupling of the Higgs field to the Ricci scalar. The value of $\xi$ is not known exactly, as will be discussed later. We shall use the largest possible $\xi\sim 10^9$ since this leads to the weakest isocurvature constraint. Then the Hubble scale \eqref{Hubble} yields $H_I \sim 2.2 \cdot 10^8 \, \text{GeV}$, in accordance with \cite{Tenkanen:2019xzn}.

For incorporating the axionic isocurvature bound \eqref{isocurvatureBoundDecayConstant} into Palatini HI, one is tempted to estimate $\sigma_\theta=H_I/(2\pi f_a)$ as in \cite{Tenkanen:2019xzn}, where $f_a$ is the axionic decay constant. With the Hubble scale \eqref{Hubble}, we would then get (for $\xi \sim 10^9$):
\begin{equation} \label{boundNaive}
	f_a \theta_i  \gtrsim 9.0 \cdot 10^{-2} \mathcal{F}^a_{\text{DM}} \frac{M_P}{\sqrt{\xi}}  \sim  \mathcal{F}^a_{\text{DM}}\,   6.9\cdot 10^{12} \, \text{GeV} \;.	
\end{equation}
If \eq \eqref{boundNaive} were to hold, plugging in \eq \eqref{thetaOff} would show that having all DM in axions, $\mathcal{F}^a_{\text{DM}}=1$, can be achieved for $f_a\sim 10^{14} \, \text{GeV}$ corresponding to $\theta_i/(2\pi) \sim 10^{-2}$ \cite{Tenkanen:2019xzn}. We shall demonstrate, however, that this conclusion is premature because it overlooks the need to canonically normalize the axion field in the early Universe.

\paragraph*{QCD axion coupled to Palatini HI}
For a fundamental derivation, the relevant action of the axion $a$ and the Higgs field $h$ (in unitary gauge) coupled to gravity is
\begin{equation} \label{actionStart}
	\begin{split}
		S= &\int \mathrm{d}^{4} x \sqrt{-g} \Big[\frac{M_P^2}{2} \Omega^2  R -  \frac{1}{2}\partial_\alpha h \partial^\alpha  h - \frac{\lambda}{4} h^4\\
		& -  \frac{1}{2}\partial_\alpha a \partial^\alpha  a - \frac{1}{2} \text{Tr} G^{\mu \nu} G_{\mu \nu}   +  \frac{a}{f_a} c_G \text{Tr} G^{\mu \nu} \tilde{G}_{\mu \nu} \Big] \;,
	\end{split}
\end{equation}
where $M_P$ denotes the reduced Planck mass, $R$ the Ricci scalar, $\lambda$ the Higgs self-coupling, and we defined
\begin{equation} \label{omega}
	\Omega^2 = 1 + \frac{\xi h^2}{M_P^2} \;,
\end{equation}
with non-minimal coupling constant $\xi$. Moreover, $G_{\mu \nu}$ and  $\tilde{G}_{\mu \nu}$ correspond to the field strength tensor of QCD and its dual, respectively, and, we have the dimensionless parameter $c_G \sim \alpha$, where $\alpha$ is the gauge coupling. We view \eq \eqref{actionStart} as the low-energy effective field theory of a pseudo-scalar $a$ coupled to QCD via an operator of mass dimension $5$ that is suppressed by a fixed mass scale $f_a$. Crucially, the shift symmetry of the axion prevents the appearance of a non-minimal interaction of $a$ with $R$.

As usual, we remove the non-minimal coupling to curvature with the conformal transformation $g_{\mu\nu} \rightarrow \Omega^{-2} g_{\mu\nu}$. In contrast to the situation in metric gravity, the Ricci tensor $R_{\mu\nu}$ of Palatini GR is independent of the metric $g_{\mu\nu}$ and therefore curvature simply transforms as $R\rightarrow \Omega^2 R$. Thus, action \eqref{actionStart} becomes
\begin{equation} \label{actionEquiv}
	\begin{split}
		S= &\int \mathrm{d}^{4} x \sqrt{-g} \Big[\frac{M_P^2}{2} \mathring{R} -  \frac{1}{2 \Omega^{2}}\partial_\alpha h \partial^\alpha  h - \frac{\lambda}{4 \Omega^{4}} h^4\\
		& -  \frac{1}{2 \Omega^{2}}\partial_\alpha a \partial^\alpha  a - \frac{1}{2} \text{Tr} G^{\mu \nu} G_{\mu \nu}   +  \frac{a}{f_a} c_G \text{Tr} G^{\mu \nu} \tilde{G}_{\mu \nu} \Big] \;,
	\end{split}
\end{equation}
where it is important to note that the gauge kinetic term is invariant under the conformal transformation so that the gauge field remains canonical (see \cite{Rigouzzo:2023sbb}).
Since in the form \eqref{actionEquiv} the coupling of gravity to matter is minimal, the Palatini and metric formulations of GR are equivalent. This has allowed us to replace curvature by its Riemannian counterpart $\mathring{R}$, which is a function of the metric only.

Still following the standard analysis of Palatini HI, we next perform a field transformation for the Higgs field, introducing a new field $\chi$ defined by \cite{Bauer:2008zj,Shaposhnikov:2020fdv}
\begin{equation} \label{canonicalHiggs}
	h = \frac{M_P}{\sqrt{\xi}} \sinh\left(\frac{\sqrt{\xi} \chi}{M_P}\right) \;,
\end{equation}
so that the action \eqref{actionEquiv} becomes
\begin{equation} \label{actionCanonical1}
	\begin{split}
		S= &\int \mathrm{d}^{4} x \sqrt{-g} \Big[\frac{M_P^2}{2} \mathring{R} -  \frac{1}{2}\partial_\alpha \chi \partial^\alpha  \chi - U-  \frac{\partial_\alpha a \partial^\alpha  a}{2 \cosh^2 \left(\frac{\sqrt{\xi} \chi}{M_P}\right)}\\
		&   - \frac{1}{2} \text{Tr} G^{\mu \nu} G_{\mu \nu} +  \frac{a}{f_a} c_G \text{Tr} G^{\mu \nu} \tilde{G}_{\mu \nu} \Big] \;,
	\end{split}
\end{equation}
with inflationary potential
\begin{equation}
	U = \frac{\lambda M_P^4}{4 \xi^2} \tanh^4 \left(\frac{\sqrt{\xi} \chi}{M_P}\right)  \;. \label{potential}
\end{equation}
Evaluating the first slow-roll parameter, we can express $\chi$ as a function of the number $N_\star$ of e-foldings (see \cite{Poisson:2023tja}) 
\begin{equation}
	\chi(N) \approx \frac{M_P \text{arccosh} \left(16 \xi N_\star\right)}{2 \sqrt{\xi}} \;, 
	\label{chi_of_N}
\end{equation}
and then match the amplitude of CMB perturbations to obtain the constraint $\xi = 1.2 \cdot 10^{10} \lambda$, where we used $N_\star\approx 51$ as in \cite{Shaposhnikov:2020fdv}. Then the potential \eqref{potential} yields the Hubble scale \eqref{Hubble}. Due to uncertainties in the measurement of the top Yukawa coupling, the inflationary value of $\lambda$ is unknown \cite{Bezrukov:2014ina,Shaposhnikov:2020fdv}. Although RG analysis indicates a small $\lambda\sim 10^{-3}$ at high energies \cite{Shaposhnikov:2020fdv}, we shall follow \cite{Tenkanen:2019xzn} and use $\lambda = 0.1$ corresponding to $\xi\sim 10^9$ since a large $\xi$ weakens isocurvature perturbations.

Finally, we define an approximately canonical axion field 
\begin{equation} \label{canonicalAxion}
	A = \frac{a}{\Omega} = \frac{a}{\cosh\left(\frac{\sqrt{\xi} \chi}{M_P}\right)} \;,
\end{equation}
where we plugged the field \eqref{canonicalHiggs} into \eq \eqref{omega}. We arrive at
\begin{equation} \label{actionCanonical2}
	\begin{split}
		S= &\int \mathrm{d}^{4} x \sqrt{-g} \Big[\frac{M_P^2}{2} \mathring{R} -  \frac{1}{2}\partial_\alpha \chi \partial^\alpha  \chi - U -  \frac{1}{2}\partial_\alpha A \partial^\alpha  A\\
		&  - \frac{1}{2} \text{Tr} G^{\mu \nu} G_{\mu \nu}  +  \frac{\Omega A}{f_a} c_G \text{Tr} G^{\mu \nu} \tilde{G}_{\mu \nu} + \Delta \mathcal{L} \Big] \;,
	\end{split}
\end{equation}
where $\Delta \mathcal{L}$ contains correction terms: 
\begin{align} \label{DeltaL}
	\Delta \mathcal{L} &= -\frac{\sqrt{\xi} A}{M_P} \tanh\left(\frac{\sqrt{\xi} \chi}{M_P}\right)\Bigg[ \partial_\alpha \chi \partial^\alpha A \nonumber\\
	& + \frac{\sqrt{\xi} A}{2 M_P} \tanh\left(\frac{\sqrt{\xi} \chi}{M_P}\right) \partial_\alpha \chi \partial^\alpha\chi \Bigg] \;.
\end{align}
Since during inflation $A\sim H_I \sim \sqrt{\lambda} M_P/\xi$, the correction terms in $\Delta \mathcal{L}$ are at least suppressed as $\sqrt{\xi} A/M_P\lesssim \sqrt{\lambda/\xi}$ and we shall neglect them in the following (see also \cite{Fairbairn:2014zta,Ballesteros:2016xej,Boucenna:2017fna,Ballesteros:2021bee} and appendix \ref{app:DeltaL} of the Supplemental Material for a consistency check).

\paragraph*{Field-dependent decay constant}
Crucially, \eqs \eqref{canonicalAxion} and \eqref{actionCanonical2} make evident that the effective decay constant $f_{a,\text{inf}}$ during inflation differs from its low-energy value $f_a$:
\begin{equation} \label{inflationaryDecayConstant}
f_{a,\text{inf}}= \frac{f_a}{\Omega} \approx \frac{f_a}{\sqrt{8 \xi N_\star}} \;,
\end{equation}
where we used the solution \eqref{chi_of_N} in \eq \eqref{omega}.
Therefore, plugging $\sigma_\theta=H_I/(2\pi f_{a,\text{inf}})$ together with \eq \eqref{Hubble} into the second condition of the isocurvature bound \eqref{isocurvatureBoundDecayConstant} implies 
\begin{equation} \label{boundFinal}
	f_a \theta_i  \gtrsim 0.25 \sqrt{N_\star} \mathcal{F}^a_{\text{DM}}  M_P\;.
\end{equation}
This constraint is much stronger than the previously proposed \eq \eqref{boundNaive}. Interestingly, it is not sensitive to the uncertainty in $\lambda$ (and equivalently $\xi$). Upon inserting $\theta_i$ as in \eq \eqref{thetaOff}, \eq \eqref{boundFinal} would bound the abundance of axions as 
\begin{equation} \label{FaDM}
	\mathcal{F}^a_{\text{DM}} \lesssim 1.1 \cdot 10^{-8} \left(\frac{	f_a}{M_P}\right)^{5/6} \;.
\end{equation}
Thus, the first condition of the isocurvature bound \eqref{isocurvatureBoundDecayConstant} is relevant for any subplanckian decay constant:
\begin{equation}
	\mathcal{F}^a_{\text{DM}} \lesssim 6.4\cdot 10^{-6} \;.
\end{equation}
Therefore, QCD axions that are present as massless field during Palatini HI can at most contribute a tiny fraction $\sim 6\cdot 10^{-6}$ to DM. We show in appendix \ref{app:UV} of the Supplemental Material that the same conclusion can be obtained in a UV-completion by a PQ-symmetry, where $a$ arises as phase of a complex PQ-field.

\paragraph*{Axion and generic inflationary models}
Clearly, the existence of these bounds does not depend on the particular structure of Palatini HI. Only important is a non-minimal coupling $\Omega^{2}\gg1$ so that the axion kinetic term is multiplied by $1/\Omega^{2}\ll1$ as in the action \eqref{actionEquiv}. This directly translates to a decay constant that is smaller during inflation than now, according to the first equality in \eq \eqref{inflationaryDecayConstant}. Importantly, the same conclusion also holds in the metric formulation of GR. In this case, the conformal transformation $g_{\mu\nu} \rightarrow \Omega^{-2} g_{\mu\nu}$ yields an additional contribution proportional to $(\partial_\alpha \Omega)^2$ (see \cite{Rigouzzo:2022yan}), but this only contributes to the kinetic term of the inflaton. Therefore, the enhancement of isocurvature perturbations is not limited to HI but generic in all inflationary models with a large non-minimal coupling to gravity.\footnote
{In a UV-completion by a PQ-field, inflationary models driven by the radial mode $|\Phi_{\text{PQ}}|$ represent an exception \cite{Fairbairn:2014zta,Ballesteros:2016xej,Boucenna:2017fna,Ballesteros:2021bee}. Since in this case $|\Phi_{\text{PQ}}|$ is displaced from its minimum, the effective inflationary decay constant can be larger than its low-energy counterpart in spite of the presence of a non-minimal coupling to gravity.}

Therefore, the strengthening of isocurvature bounds that we have identified applies to leading inflationary plateau models, in particular Starobinsky inflation \cite{Starobinsky:1980te} (see appendix \ref{app_starobinsky} for details), metric HI \cite{Bezrukov:2007ep}, and certain classes of attractor models \cite{Kallosh:2013tua,Galante:2014ifa}. Even without the effect of the conformal transformation, these scenarios are typically incompatible with isocurvature bounds because of their high inflationary Hubble scale. However, any attempt to alleviate this must additionally account for the further suppression of the inflationary decay constant induced by the non-minimal coupling.\footnote
{In the recent paper \cite{Graham:2025iwx}, which appeared after our work, a mechanism was constructed for alleviating isocurvature constraints based on a non-minimal coupling $\xi_{\text{PQ}}$ of the radial PQ field to curvature. They discuss isocurvature constraints in Starobinsky inflation, but without accounting for the modification of the axion kinetic term induced by the conformal transformation. In \cite{toAppear}, we study the combined effect of both non-minimal couplings $\xi$ and $\xi_{\text{PQ}}$, showing that the inflaton coupling $\xi$ significantly reduces the viable interval of $\xi_{\text{PQ}}$ in which isocurvature bounds can be alleviated.}
We note that $\alpha$-attractor models represent an exception. While they can be constructed without invoking a non-minimal coupling \cite{Kallosh:2013hoa,Kallosh:2013yoa}, some can also be obtained from a negative non-minimal coupling, $\xi<0$ \cite{Kallosh:2013hoa,Kallosh:2013maa}. In this case, $\Omega<1$, which could open up the possibility of alleviating isocurvature bounds.

Conversely, the non-minimal interaction can lead to new channels for detecting axions in those models that are not ruled out: We expect that the rapid change of $\Omega$ during reheating leads to the production of axions (\cf \cite{Ema:2016dny}), which may result in an observable signal in the effective number of massless degrees of freedom \cite{Baumann:2016wac} (see also \cite{DiLuzio:2020wdo}).

\paragraph*{Way out from non-minimal coupling to torsion}
Since Palatini gravity is part of the Einstein-Cartan formulation (see \cite{Rigouzzo:2022yan} for terminology), we can add many more terms composed of torsion to the action (\cf \cite{Karananas:2021zkl,Rigouzzo:2022yan,Rigouzzo:2023sbb}). All of them could potentially modify the value of the decay constant. As an illustrative example, we shall consider a direct coupling of the axion to torsion via a term $\zeta J_\alpha T^\alpha$, where $\zeta$ is a coupling constant, $J_\alpha=f_a \partial_\alpha a$, and $T^\alpha=g_{\mu \nu} T^{\mu \alpha \nu}$ is the torsion vector (with $T^\mu_{ \ \alpha \nu}$ defined in terms of the Christoffel symbols $\Gamma^\mu_{ \ \alpha \nu}$ as  $T^\mu_{ \ \alpha \nu}\equiv 1/2(\Gamma^\mu_{ \ \alpha \nu} - \Gamma^\mu_{ \  \nu \alpha}$). Correspondingly, action \eqref{actionStart} is extended as
\begin{equation}
	\begin{split}
		S= &\int \mathrm{d}^{4} x \sqrt{-g} \Big[\frac{M_P^2}{2} \Omega^2  R -  \frac{1}{2}\partial_\alpha h \partial^\alpha  h - \frac{\lambda}{4} h^4\\
		& -  \frac{1}{2}\partial_\alpha a \partial^\alpha  a - \frac{1}{2} \text{Tr} G^{\mu \nu} G_{\mu \nu}   +  \frac{a}{f_a} c_G \text{Tr} G^{\mu \nu} \tilde{G}_{\mu \nu} \\& -\zeta J_\alpha T^\alpha \Big]\;.
	\end{split}
	\label{way_out_init}
\end{equation}
After solving for $T^\alpha$ and going to the Einstein frame via the conformal transformation $g_{\mu\nu} \rightarrow \Omega^{-2} g_{\mu\nu}$, we derive an extension of \eq \eqref{actionEquiv} (see details in appendix \ref{app:wayOut} of the Supplemental Material). From it we can read off the non-trivial axionic kinetic term:
\begin{equation} \label{kineticTermWayOut}
	-  \frac{1}{2\Omega^2}\left(1-\frac{3 \zeta^2 f_a^2}{2 M_P^2 \Omega^2} \right)\partial_\alpha a \partial^\alpha  a \;.
\end{equation}
Restricting ourselves to the late Universe, where $\Omega^2 \sim 1$, we canonically normalize $a$ to show that the decay constant is modified at low energies:
\begin{equation} 
	f_a  \rightarrow f_{a,\text{IR}}(f_a) = \sqrt{1-\frac{3 \zeta^2 f_a^2}{2 M_P^2}} f_a \;.
	\label{f_a_ir}
\end{equation}
Now $f_{a,\text{IR}}$ becomes the scale that suppresses the operator $a\, c_G \text{Tr} G^{\mu \nu} \tilde{G}_{\mu \nu}$, which among others is responsible for the non-perturbative generation of the axion mass in the late Universe.\footnote
{In a UV-completion with PQ-symmetry, $J_\alpha=f_a \partial_\alpha a$ arises from the PQ-current $J_\alpha\equiv -i\left(\Phi_{PQ}^\star \partial_\alpha \Phi_{PQ} - (\partial^\alpha \Phi_{PQ})^\star \Phi_{PQ} \right)$ after PQ symmetry breaking. Therefore, $f_a$ still sets the expectation value of the canonical PQ-field at low energies.}
Therefore, \eq \eqref{boundFinal} remains valid, with the only difference that $\theta_i$ (and hence also $\mathcal{F}^a_{\text{DM}}$) now depend on $f_{a,\text{IR}}$. Thus, the bound \eqref{FaDM} generalizes to  (\cf \cite{Ballesteros:2021bee})
\begin{equation}
	\mathcal{F}^a_{\text{DM}} \lesssim 	1.1 \cdot 10^{-8} \left( \frac{f_{a,\text{IR}}}{M_P}\right)^{-7/6}	\left( \frac{f_{a}}{M_P} \right)^2 \;.
\end{equation}
After plugging in the low-energy decay constant \eqref{f_a_ir}, we conclude that axions can constitute all of DM if 
\begin{equation} \label{parametersOut}
	1-2 \cdot 10^{-14} \left( \frac{f_a}{M_P}\right)^{10/7} \lesssim \frac{3 \zeta^2 f_a^2}{2 M_P^2} < 1 \;.
\end{equation}

\begin{figure}
	\centering
	\includegraphics[width= 0.8\linewidth]{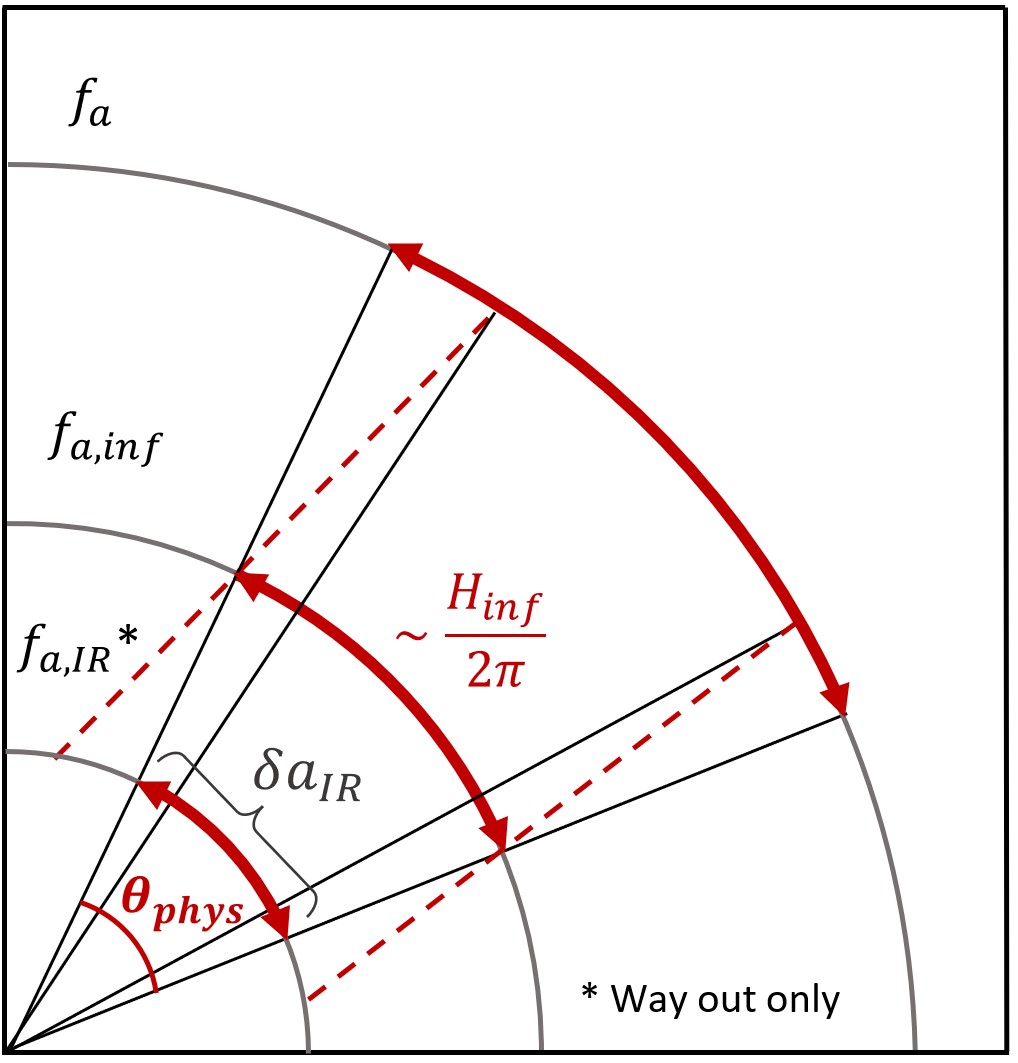}
	\caption{Representation of a possible way out: If the effective decay constant becomes smaller at lower energy, this improves the isocurvature bound and allows for the axion to account for all of DM. The \textit{physical} fluctuation of the axion field at low energy is shown with $\delta a_{IR}$. Figure inspired by \cite{Fairbairn:2014zta}.}
	\label{diagram2}
\end{figure}

This implies $f_{a,\text{IR}}\lesssim 10^{-7} f_a$ and so comparison with \eq \eqref{inflationaryDecayConstant} shows that we get the hierarchy $f_{a,\text{IR}}<f_{a,\text{inf}}<f_a$. A decay constant that is larger during inflation than in the late Universe alleviates isocurvature bounds, as in \cite{Fairbairn:2014zta,Ballesteros:2016xej,Boucenna:2017fna,Ballesteros:2021bee} -- see visual depiction of the mechanism in \fig \ref{diagram2}.
The value \eqref{parametersOut} does not change the high-energy decay constant $f_{a,\text{inf}}$ since \eq \eqref{kineticTermWayOut} shows that during inflation the correction resulting from $\zeta$ is negligible, which makes our using of \eqs \eqref{inflationaryDecayConstant} and \eqref{boundFinal} self-consistent.
Evidently, one could include numerous other torsion contributions in the action, and a systematic study of their effect remains to be performed \cite{longPaper}. 

\paragraph*{Further ways out} Several other options exist to alleviate isocurvature bounds. First, it is possible to consider non-minimal models of HI such as \cite{Langvik:2020nrs,Shaposhnikov:2020gts} or to include a direct coupling between the norm of the PQ-field and torsion, $|\Phi_{\text{PQ}}|^2 T_\alpha T^\alpha$. Independently of torsion, one can consider an early phase of confinement \cite{Dvali:1995ce}, which can arise from an effective coupling of the Higgs field to the gauge kinetic term $\text{Tr} G^{\mu \nu} G_{\mu \nu}$ and would drive the QCD axion to the CP-conserving value already during inflation, thereby reducing isocurvature perturbations. We do not expect this mechanism to be generic in our approach since there is no simple coupling of torsion to gauge fields and so new new interactions involving $G^{\mu \nu}$ arise \cite{Rigouzzo:2023sbb}. In a UV-completion by a PQ-field, one can also directly include a coupling $|\Phi_{\text{PQ}}|^2 h^2$ to change the inflationary vacuum expectation value of the PQ-field (see also \cite{Nakayama:2015pba}),\footnote{We thank Misha Shaposhnikov for pointing out this option.} or even allow for an explicit breaking of PQ-symmetry \cite{Dine:2004cq,Higaki:2014ooa,Takahashi:2015waa,Barenboim:2024akt,Berbig:2024ufe}. However, all of these options require additional non-minimal interactions with coupling constants in specific intervals -- \cf the highly tuned parameter choice \eqref{parametersOut} -- and so arguably should be considered as non-minimal. Finally, one may turn to more generic axion-like particles (see \cite{Marsh:2015xka}) as solution to the strong CP-problem, as \eg in \cite{Karananas:2024xja}.

\paragraph*{Conclusion} The microscopic origins of inflation and dark matter remain among the most important open questions in cosmology. Given that many models are still consistent with all observations, additional constraints are needed to distinguish among the plethora of proposals. In this work, we have corrected a previous claim in the literature and demonstrated that a massless QCD axion -- if abundant enough to account for dark matter in the late Universe -- must not exist during simple Higgs inflation models. This finding resonates with  the approach of Higgs inflation, which is motivated by the lack of detected particles beyond the Standard Model.

Our result stems from the energy dependence of the axionic decay constant. An inflationary value that is smaller than its late-time counterpart enhances isocurvature perturbations. As a result, a significant axion abundance is incompatible with observational isocurvature bounds. Our findings have important implications for all inflationary models with non-minimal couplings to gravity. First, the enhancement of isocurvature perturbations is a general feature of such scenarios. Second, coupling the axion to torsion can relax isocurvature constraints; in particular, Einstein-Cartan gravity can rescue proposal otherwise deemed excluded due to excessive isocurvature perturbations.

\begin{acknowledgments}

\paragraph*{Acknowledgments} We thank Maximilian Berbig, Gia Dvali, Mudit Jain, Georgios Karananas, David Marsh, Nicole Righi, and Misha Shaposhnikov for discussions and
	insightful feedback. C.R.~is very grateful to Mudit Jain for long and enlightening discussions on axion abundance and isocurvature perturbations. C.R.~acknowledges support from the
Science and Technology Facilities Council (STFC). The work of S.Z.~was supported by the European Research Council Gravites Horizon Grant AO number: 850 173-6.

\textbf{Disclaimer:} Funded by the European Union. Views and opinions expressed are however those of
the authors only and do not necessarily reflect those of the European Union or European Research
Council. Neither the European Union nor the granting authority can be held responsible for them.
\end{acknowledgments}

\appendix
\section{Abundance of the QCD axion} \label{app:abundance}
The potential of the axion for small $\theta$ reads: 
\begin{equation}
	V(\theta) \simeq \frac{1}{2} m_a^2f_a^2 \theta^2 \;.
\end{equation}
When $ H\gg m/3$, the axion freezes (the Hubble friction prevents if from rolling) and is approximately given by $\theta \sim \theta_i=\text{constant}$. At this stage, the number density is frozen and given by 
\begin{equation}
	n_i=\frac{\rho}{m_a}=\frac{V(\theta_i)}{m_a}=\frac{1}{2}m_a f_a^2 \theta_i^2 \;,
\end{equation}
which holds for non-relativistic particles. When $H \lesssim m_a/3$, $\theta$ starts oscillating, therefore becoming temperature dependent: $\theta_i \rightarrow \theta(T)$. The number density then becomes: 
\begin{equation}
	n(T)= \frac{1}{2}m(T) \theta^2(T) f_a^2 \;.
\end{equation}
We also know that during matter domination epoch
\begin{equation}
	n(T)=\frac{n_i a^3(T_i)}{a^3(T)}=\frac{1}{2}m(T_i) \theta_i^2  f_a^2 \left( \frac{g_s(T)}{g_s(T_i)}\right)\frac{T^3}{T_i^3} \;.
	\label{number_density}
\end{equation}
Finally, the QCD axion mass depends on temperature, which can be approximated as \cite{Borsanyi:2016ksw}:
\begin{equation}
	m(T) \sim \underbrace{5.7 \times 10^{-10} \text{eV}\left( \frac{10^{16} \text{GeV}}{f_a} \right)}_{m_a} \begin{cases}
		1 , \quad  &T < T_c \\
		(T_c/T)^4 ,\quad &T > T_c
	\end{cases} \;,
\end{equation}
with $T_c \sim 150 $ MeV.
Therefore we can now solve for $T_i$: 
\begin{equation}
	\begin{aligned}
		&3H(T_i)=m(T_i) \;,\\ 
		& \Leftrightarrow 3 \sqrt{\frac{\pi^2}{90}g_p(T_i)}\frac{T_i^2}{M_{P}}=m_a \left( \frac{T_c}{T_i}\right)^4 \;,\\
		&\Rightarrow T_i^6=\frac{m_a M_{P}}{3\sqrt{\frac{\pi^2}{90}g_p(T_i)}}T_c^4 \;.
	\end{aligned}
\end{equation}
We can also express the mass of the QCD axion at freeze out: 
\begin{equation}
	m(T_i)=m_a \frac{T_c^4}{\left(\frac{m_a M_P}{3\sqrt{\frac{\pi^2}{90}g_p(T_i)}}T_c^4 \right)^{4/6}} \;.
\end{equation}
Plugging this into \eq \eqref{number_density}, we obtain
\begin{equation}
	n(T)= \frac{1}{2} m_a \frac{T_c^4}{\left(\frac{m_a M_P}{3\sqrt{\frac{\pi^2}{90}g_p(T_i)}}T_c^4 \right)^{4/6}} \theta_i^2 f_a^2 \frac{g_s(T)}{g_s(T_i)} \frac{T^3}{T_i^3} \;.
\end{equation}
Finally, the axion energy density nowadays is given by $\rho_a=m_a n(T)$ because we are at a temperature $T<T_C$. So: 
\begin{equation}
	\begin{aligned}
		\rho_a(T)&=\frac{1}{2}m_a^2 f_a^2 \frac{g_s(T)}{g_s(T_i)} T_c^{4} T^3 \left(\frac{m_a M_{P}}{3\sqrt{\frac{\pi^2}{90}g_p(T_i)}}T_c^4\right)^{-7/6} \theta_i^2\\
		&= \frac{1}{2}m_a^2 f_a^2 \frac{g_s(T)}{g_s(T_i)} T_c^{-4/6} T^3 \left(
		\frac{3 f_a\sqrt{\frac{\pi^2}{90}g_p(T_i)}}{5.7 \times 10^{15} M_{P}}\right)^{7/6} \theta_i^2 \;.
	\end{aligned}
\end{equation}
From there, we can get the abundance of axion \cite{Sheridan:2024vtt} 
\begin{equation} \label{abundance}
	\Omega_{a}h^2\sim 0.12 \left( \frac{\theta_{i}}{4.7 \times 10^{-3}}\right)^2 \left(\frac{f_a}{10^{16}\text{GeV}}\right)^{7/6} \;,
\end{equation}
where we used $M_P = 2.43 \times 10^{27}$ eV, 
$T_c = 150 \times 10^{6}$ eV, 
$T_0 = 2.33*10^{-4}$ eV, $\rho_c=8.06*10^{-11} h^2$ eV,
$g_s(T_0) = 4$ \cite{Saikawa:2018rcs},
$g_s(T_i) = 41$, and $g_p(T_i)=44$ \cite{Sheridan:2024vtt}.
\\If the axion was all of DM nowadays, then the \lhs would be $0.12$. Now, if the axion is a fraction of the DM, then: 
\begin{equation} \label{FaApp}
	\mathcal{F}^a_{\text{DM}} \sim \left( \frac{\theta_{i}}{4.7 \times 10^{-3}}\right)^2 \left(\frac{f_a}{10^{16}\text{GeV}}\right)^{7/6} \;,
\end{equation}
or equivalently
\begin{equation} \label{faOfTheta_app}
	f_a = \mathcal{F}^{a\, 6/7}_{\text{DM}} f_0 \,	 \theta_{i}^{-12/7} \;, 
\end{equation}
where we defined\footnote
{In \cite{Tenkanen:2019xzn} a smaller value $f_0 = 1.5\cdot 10^{11}\, \text{GeV}$ was used. Since the bound on the axion abundance \eqref{FaDM} scales with a positive power of $f_0$, \eq \eqref{f0} yields the more conservative result. The small difference could be due to a) a different value for $T_C$ (in \cite{DiLuzio:2020wdo} $T_C=160$ MeV is used), b) a different value for the degrees of freedom (in \cite{DiLuzio:2020wdo} $g_p(T_i)=g_s(T_i) \sim 61.75$ is used), c) the use of a refined formula for the axion mass at high energies $m_a(T) \simeq \beta m_a\left(\frac{T_C}{T}\right)^4$, with $\beta$ a parameter that depends on the quark flavors physics, that \cite{DiLuzio:2020wdo} and \cite{RevModPhys.53.43} estimates it at $10^{-2}$.}
\begin{equation} \label{f0}
	f_0 = 1.02 \times 10^{12} \text{GeV} \;.
\end{equation}
Solving \eq \eqref{faOfTheta_app} for $\theta_i$, we get \eq \eqref{thetaOff} as shown in the main part.
\section{Axion isocurvature bounds} \label{app:isocurvature}
As shown in \eq \eqref{isocurvatureAxion}, the general formula for the isocurvature perturbation is \cite{Beltran:2006sq,Hertzberg:2008wr}: 
\begin{equation}
	\Delta_a=\mathcal{F}^a_{\text{DM}} \frac{\sigma_\theta \sqrt{2(\sigma^2_\theta+2 \theta_i^2)}}{\theta_i^2+\sigma_\theta^2} \;.
\end{equation}
Without any assumptions, we can bound
\begin{equation}
		\Delta_a<\mathcal{F}^a_{\text{DM}} \frac{2\sigma_\theta }{\sqrt{\theta_i^2+\sigma_\theta^2}}< \mathcal{F}^a_{\text{DM}} \frac{2 \sigma_\theta}{\theta_i} \;.
\end{equation}
 Still without restrictions, one can alternatively bound
\begin{equation}
	\Delta_a<\mathcal{F}^a_{\text{DM}} \frac{2\sigma_\theta }{\sqrt{\theta_i^2+\sigma_\theta^2}}< 2 \mathcal{F}^a_{\text{DM}} \;.
\end{equation}
For obtaining lower bounds, we distinguish two cases. If $\sigma_\theta<\theta_i$, then
\begin{equation}
		\Delta_a>\mathcal{F}^a_{\text{DM}} \frac{\sqrt{2}\sigma_\theta }{\sqrt{\theta_i^2+\sigma_\theta^2}} > \mathcal{F}^a_{\text{DM}} \frac{ \sigma_\theta}{\theta_i} \;.
\end{equation}
In the other case,  $\sigma_\theta>\theta_i$, we get 
\begin{equation}
	\Delta_a>\mathcal{F}^a_{\text{DM}} \frac{\sqrt{2}\sigma_\theta }{\sqrt{\theta_i^2+\sigma_\theta^2}} > \mathcal{F}^a_{\text{DM}} \;.
\end{equation}

In summary, we conclude:
\begin{align}
\sigma_\theta<\theta_i:& \qquad 	 \mathcal{F}^a_{\text{DM}} \frac{ \sigma_\theta}{\theta_i} < 	\Delta_a < 2  \mathcal{F}^a_{\text{DM}} \frac{\sigma_\theta}{\theta_i}\;,\\
\sigma_\theta>\theta_i:& \qquad \mathcal{F}^a_{\text{DM}} < 	\Delta_a < 2 \mathcal{F}^a_{\text{DM}} \;.
\end{align}
Thus, the asymptotic scalings $\Delta_a \approx \mathcal{F}^a_{\text{DM}} 2 \sigma_\theta/\theta_i$ (for $\sigma_\theta \ll \theta_i$) and $\Delta_a\approx \sqrt{2} \mathcal{F}^a_{\text{DM}}$ (for $\sigma_\theta \gg \theta_i$) represent good approximations, with an error of at most $2$, even when $\sigma_\theta$ and $\theta_i$ are of the same order of magnitude. Therefore, the requirement \eqref{isocurvatureBoundDecayConstant} is applicable in the full parameter space of $\sigma_\theta$ and $\theta_i$. 

In the most general model, with three isocurvature parameters, CDI Planck TT,TE,EE+lowE+lensing gives the upper limit on $ \beta = \Delta^2_a/(\Delta^2_{\mathcal{R}}+\Delta_a^2)$ between $0.01$ and $0.47$, and $-0.12 <\cos (\Delta)<0.15$ \cite{Planck:2018jri}. Let us be the most conservative possible and use $\beta <0.47$. Propagating this onto the isocurvature bounds equations we find: 
\begin{equation} 
	\mathcal{F}^a_{\text{DM}} \lesssim 3.1\cdot 10^{-5} \quad\ \text{OR}\ \quad	\sigma_\theta \lesssim 2.2 \cdot  10^{-5} \frac{\theta_i}{\mathcal{F}^a_{\text{DM}}} \;.
\end{equation}
Therefore, our conclusion remain: QCD axions present as massless field during Palatini HI can at most contribute to a tiny fraction $\sim 10^{-5}$ to DM.
\section{On Mixing of inflaton and axion} \label{app:DeltaL}
In order to remove the leading mixing term in \eq \eqref{DeltaL}, we can perform another field redefinition
\begin{equation}\label{anotherFieldTrafo}
	\chi \rightarrow \chi -1/2 \frac{\sqrt{\xi} A^2}{M_P} \tanh\left(\frac{\sqrt{\xi} \chi}{M_P}\right) .
\end{equation}
Apart from terms that are suppressed by at least two powers of $\sqrt{\xi}A/M_P$, this generates a potential for the axion. In order to evaluate it, we use the well-known approximation of the potential \eqref{potential} (see \cite{Shaposhnikov:2020gts})
\begin{equation}
	U = \frac{\lambda M_P^4}{4\xi^2}\left(1 + \exp\left\{-\frac{2\sqrt{\xi} \chi}{M_P}\right\}\right)^{-2} .
\end{equation}
Plugging the field transformation \eqref{anotherFieldTrafo} into this asymptotic form of the potential, we expand to second order in $A$: 
\begin{align}
	U \approx& \frac{\lambda M_P^4}{4\xi^2}\left(1 + \exp\left\{-\frac{2\sqrt{\xi} \chi}{M_P}\right\}\right)^{-2}\nonumber\\
& -\frac{ \lambda A^2  M_P^2}{2 \xi} e^{-\frac{2\sqrt{\xi} \chi}{M_P}}  \left(1+ e^{-\frac{2\sqrt{\xi} \chi}{M_P}} \right)^{-3}  \;.
\end{align}

We see that the induced mass of $A$ scales as
\begin{equation}
|m_A^2|\sim	\frac{\lambda  M_P^2 }{\xi \exp \left(\frac{2\sqrt{\xi} \chi}{M_P}\right)} \;.
\end{equation}
Since $\exp \left(\sqrt{\xi} \chi/M_P\right)\sim \sqrt{\xi N_\star}$ by \eq \eqref{chi_of_N}, we conclude that the mass is suppressed, $|m_A|\sim \sqrt{\lambda}\dfrac{M_P}{\xi \sqrt{N_\star}}$ and in particular $m_A <H_I\sim \sqrt{\lambda}\dfrac{M_P}{\xi}$. In summary, we can remove the leading correction term from \eq \eqref{DeltaL}, which only produces a negligible axion mass. This is a consistency check showing that indeed the suppressed terms in \eq \eqref{DeltaL} can be safely neglected.

	\section{Consistency with UV-perspective} \label{app:UV}
In the main part, we have employed an effective field theory approach in which we treat the axion $a$ as a low-energy degree of freedom, independently of its UV-completion (see \cite{Witten:1984dg,Reece:2024wrn} and \cite{Dvali:2005an,Dvali:2013cpa,Dvali:2017mpy,Dvali:2022fdv} for alternatives to the original proposal of a PQ-axion \cite{Peccei:1977hh,Weinberg:1977ma,Wilczek:1977pj}). 
As a consistency check, we shall now demonstrate that our analysis is compatible with the PQ-mechanism, where $a$ arises as phase field of the complex PQ-field $\Phi_{\text{PQ}}$.
Then in \eq \eqref{actionStart} the term $1/2\, \partial_\alpha a \partial^\alpha  a$ is replaced by $|\partial_\alpha \Phi_{\text{PQ}} \partial^\alpha \Phi_{\text{PQ}}|$, which after the conformal transformation becomes $1/\Omega^2 |\partial_\alpha \Phi_{\text{PQ}} \partial^{\alpha} \Phi_{\text{PQ}}|$, to be inserted into \eq \eqref{actionEquiv}. Thus, PQ-symmetry breaking leading to $\braket{\Phi_{\text{PQ}}}=f_a/\sqrt{2}$ is equivalent to 
\begin{equation} \label{canonicalPQ}
	\braket{\tilde{\Phi}_{\text{PQ}}}= \frac{f_a}{\sqrt{2} \Omega} \,,\ \text{with}\ \tilde{\Phi}_{\text{PQ}} \equiv  \frac{\Phi_{\text{PQ}}}{\Omega}  \,,
\end{equation}
where $\tilde{\Phi}_{\text{PQ}}$ is defined to be approximately canonical (\cf \eq \eqref{canonicalAxion}). Importantly, $f_a$ is now defined as the energy scale setting the expectation value of the bare PQ-field (\eg resulting from a potential of the form $(|\Phi_{\text{PQ}}|^2 - f_a^2)^2$). Reading off the inflationary decay constant as vacuum expectation value of the canonical $\tilde{\Phi}_{\text{PQ}}$, we conclude $f_{a,\text{inf}}= f_a/\Omega$, in accordance with \eq \eqref{inflationaryDecayConstant}.

As a second point, \eq \eqref{canonicalPQ} makes clear that 
\begin{equation}
	\text{e}^{ia/f_{a}} \propto \Phi_{\text{PQ}} \propto  \tilde{\Phi}_{\text{PQ}} \propto 	\text{e}^{iA/f_{a,\text{inf}}} \,.
\end{equation}
Thus, an evolving vacuum expectation value of the PQ-field conserves the phase $a/f_a=A/f_{a,\text{inf}}$. This fact is well-known \cite{Fairbairn:2014zta} (see also \cite{Ballesteros:2016xej,Boucenna:2017fna,Ballesteros:2021bee}) and justifies our plugging in of $\sigma_\theta$ as computed during inflation into the post-inflationary constraint \eqref{isocurvatureBoundDecayConstant}.

\section{Way out from direct coupling of torsion to matter} \label{app:wayOut}
Starting from \eq \eqref{way_out_init}:
\begin{equation}
	\begin{split}
		S= &\int \mathrm{d}^{4} x \sqrt{-g} \Big[\frac{M_P^2}{2} \Omega^2  R -  \frac{1}{2}\partial_\alpha h \partial^\alpha  h - \frac{\lambda}{4} h^4\\
		& -  \frac{1}{2}\partial_\alpha a \partial^\alpha  a - \frac{1}{2} \text{Tr} G^{\mu \nu} G_{\mu \nu}   +  \frac{a}{f_a} c_G \text{Tr} G^{\mu \nu} \tilde{G}_{\mu \nu} \\& - \zeta J_\alpha T^\alpha \Big]\;,
	\end{split}
	\label{way_out_init_app}
\end{equation}
one can use that the scalar curvature depends on the metric and torsion via $
R=\mathring{R}+ 2 \mathring{\nabla}_\alpha T^\alpha - \frac{2}{3} T_\alpha T^\alpha + \frac{1}{24} \hat{T}_\alpha \hat{T}^\alpha$, 
therefore splitting the torsion part and the metric part in our initial action:
\begin{equation}
	\begin{split}
		S= & \int \mathrm{d}^{4} x \sqrt{-g} \Big[\frac{M_P^2}{2} \Omega^2  \mathring{R} -  \frac{1}{2}\partial_\alpha h \partial^\alpha  h - \frac{\lambda}{4} h^4\\
		& -  \frac{1}{2}\partial_\alpha a \partial^\alpha  a - \frac{1}{2} \text{Tr} G^{\mu \nu} G_{\mu \nu}   +  \frac{a}{f_a} c_G \text{Tr} G^{\mu \nu} \tilde{G}_{\mu \nu} \\& - \zeta J_\alpha T^\alpha + M_P^2 \Omega^2 \mathring{\nabla}_\alpha T^\alpha -\frac{M_P^2 \Omega^2}{3} T_\alpha T^\alpha +\frac{M_P^2 \Omega^2}{48} \hat{T}_\alpha \hat{T}^\alpha \Big]				
		\;.
	\end{split}
	\label{way_out_2}
\end{equation}
Then we can solve for $T^\alpha$ and $\hat{T}^\alpha$:
\begin{equation}
	T_\alpha= -\frac{3}{2 M_P^2 \Omega^2} \left(M_P^2 \partial_\alpha(\Omega^2) + \zeta J_\alpha\right) \;, \quad \hat{T}_\alpha=0 \;.
\end{equation}
Plugging it back in \eq \eqref{way_out_2} gives: 
\begin{equation}
	\begin{split}
		S= & \int \mathrm{d}^{4} x \sqrt{-g} \Big[\frac{M_P^2}{2} \Omega^2  \mathring{R} -  \frac{1}{2}\partial_\alpha h \partial^\alpha  h - \frac{\lambda}{4} h^4\\
		& -  \frac{1}{2}\partial_\alpha a \partial^\alpha  a - \frac{1}{2} \text{Tr} G^{\mu \nu} G_{\mu \nu}   +  \frac{a}{f_a} c_G \text{Tr} G^{\mu \nu} \tilde{G}_{\mu \nu} \\& +\frac{3 \zeta^2}{4 M_P^2 \Omega^2}J_\alpha J^\alpha + 3 M_P^2 \partial_\alpha\Omega\partial^\alpha\Omega + \frac{3 \zeta}{ 2 \Omega^2} \partial_\alpha(\Omega^2) J^\alpha   \Big]				
		\;.
	\end{split}
\end{equation}
Finally, we can do a conformal transformation to arrive at
\begin{equation}
	\begin{split}
		S= &\int \mathrm{d}^{4} x \sqrt{-g} \Big[\frac{M_P^2}{2}\mathring{R} -  \frac{1}{2 \Omega^{2}}\partial_\alpha h \partial^\alpha  h - \frac{\lambda}{4 \Omega^{4}} h^4 \\
		& -  \frac{1}{2 \Omega^{2}}\partial_\alpha a \partial^\alpha  a - \frac{1}{2} \text{Tr} G^{\mu \nu} G_{\mu \nu}   +  \frac{a}{f_a} c_G \text{Tr} G^{\mu \nu} \tilde{G}_{\mu \nu} \\ & +\frac{3 \zeta^2}{4 M_P^2 \Omega^4}J_\alpha J^\alpha + \frac{3 \zeta}{2 \Omega^4} \partial_\alpha(\Omega^2) J^\alpha \Big] \;,
	\end{split}
\end{equation} 
where the inhomogeneous part of the conformal transformation cancels out with $3 M_P^2 \partial_\alpha\Omega\partial^\alpha\Omega$. Plugging in $J_\alpha=f_a \partial_\alpha a$, we can read off the axionic kinetic term \eqref{kineticTermWayOut} as shown in the main part.

\section{Starobinksy inflation}
\label{app_starobinsky}
Starobinsky inflation arises from modifying the Einstein-Hilbert action with an additional quadratic curvature term:
\begin{equation}
	S = \frac{M_P^2}{2} \int d^4x \sqrt{-g} \left( R + \frac{1}{6M^2} R^2 \right) \;,
\end{equation}
where $M_P$ is the reduced Planck mass and $M$ is a mass scale related to inflation. To linearize the $R^2$ term, we introduce an auxiliary scalar field $\phi$ (see \eg \cite{Mantziris:2022fuu}):
\begin{equation}
	S = \frac{M_P^2}{2} \int d^4x \sqrt{-g} \left[ \left(1 + \frac{\phi^2}{3M^2} \right) R - \frac{\phi^4}{6M^2} \right]\;.
\end{equation}
The equation of motion for $\phi$ yields $\phi^2 = R$, recovering the original action. We then perform a conformal transformation to the Einstein frame:
\begin{equation}
	\tilde{g}_{\mu\nu} = \Omega^2 g_{\mu\nu}, \quad \text{with} \quad \Omega^2 = 1 + \frac{\phi^2}{3M^2}\;.
	\label{conformal_transformation}
\end{equation}
This conformal transformation will introduce a non-canonical kinetic term for $\phi$, because of the scalar curvature transforming non-homogeneously (see e.g \cite{Rigouzzo:2022yan}):
\begin{equation} \label{transformationR}
	R \rightarrow \Omega^2\left[R+3 \Box \ln \Omega^2-\frac{3}{2} g^{\mu \nu}\left(\partial_\mu \ln \Omega^2\right)\left(\partial_\nu \ln \Omega^2\right)\right] \;.
	\end{equation}
The action becomes:
\begin{equation}
	S = \frac{M_P^2}{2} \int d^4x \sqrt{-g} \left( R - \frac{\phi^4}{6M^2 \Omega^4} - \frac{2 \phi^2}{3 M^4 \Omega^4} \partial_\mu \phi \partial^\mu \phi \right)\;.
	\end{equation}
Introducing a canonically normalized scalar field $\chi$ via:
\begin{equation}
	\chi = \sqrt{\frac{3}{2}} M_P \ln\left(1 + \frac{\phi^2}{3M^2} \right)\;,
\end{equation}
we can express $\phi$ as:
\begin{equation}
	\phi^2 = 3M^2 \left( e^{\sqrt{\frac{2}{3}} \chi / M_P} - 1 \right)\;.
	\label{starobinksy_transfo}
\end{equation}
Substituting this into the potential term:
\begin{equation}
	V(\phi) = \frac{M_P^2}{2} \cdot \frac{\phi^4}{6M^2 \Omega^4}\;,
\end{equation}
we obtain the inflaton potential:
\begin{equation}
	V(\chi) = \frac{3}{4} M_P^2 M^2 \left( 1 - e^{-\sqrt{\frac{2}{3}} \chi / M_P} \right)^2\;.
	\label{starobinsky_potential}
\end{equation}
 We can then compute the field value when inflation ends, $\chi_\text{end}$ by finding when $\epsilon=\frac{M_P^2}{2}(\frac{V'(\chi)}{V(\chi)})^2 \simeq 1$. The slow roll parameter $\epsilon$ for the potential in Eq.(\ref{starobinsky_potential}) is:
\begin{equation}
		\epsilon = \frac{4(e^{-2\sqrt{\frac{2}{3}} \chi / M_P})}{3(1-e^{-\sqrt{\frac{2}{3}} \chi / M_P})^2} \;,
\end{equation}
therefore $\chi_\text{end}=M_P \sqrt{\frac{3}{2}} \ln\left((2 \sqrt{3} - 3)^{-1}\right) \simeq 0.94 M_P$. The number of e-folds is given by: 
\begin{equation}
	N \simeq \frac{3}{4} \exp \left(\sqrt{\frac{2}{3}} \frac{\chi_\text{in}}{M_P}\right) \;,
\end{equation}
from which we can read the generation of perturbations observed in the CMB: $\chi \sim M_P \sqrt{\frac{3}{2}}\ln(4N/3)$ This gives a conformal factor $\Omega^2 \simeq 4 N/3 \simeq 68$ for $N \simeq 51$. Finally, matching the observed amplitude of CMB perturbations \cite{Planck:2018jri}, $V/\epsilon \approx 5 \cdot 10^{-7}M_P^4$, fixes $M\approx 1.4 \cdot 10^{-5} M_P$.

Crucially, the conformal transformation implies that, if a pseudo-scalar field like an axion was present during inflation, its decay constant will be modified to
 \begin{equation}
 f_{a,\text{inf}}=\frac{f_a}{\Omega}\;.
 \end{equation} 
 Since $\Omega>1$ for Starobinsky inflation, isocurvature bounds are generically worsened. We discuss this in detail in \cite{toAppear}.

\bibliography{Refs}

\end{document}